\documentclass[prd,reprint, nofootinbib,,superscriptaddress]{revtex4-1}

\usepackage{graphicx}
\usepackage[colorlinks=true,linkcolor=blue,citecolor=red,urlcolor=blue]{hyperref}
\usepackage{amsmath, amssymb}
\usepackage{bm}
\usepackage{natbib}
\usepackage{subfigure}
\usepackage{color}
\usepackage{physics}
\usepackage{soul,xcolor}

\def\mbf{\mathbf}
\def\i{\text{i}}
\newcommand{\be}{\begin{equation}}
\newcommand{\ee}{\end{equation}}

\begin{document}
\title{Superradiant scattering in fluids of light} 
\author{Angus Prain}
\email{Angus.Prain@royalsociety.org.nz}
\affiliation{Royal Society Te Ap\={a}rangi, 11 Turnbull Street, Wellington 6011, New Zealand}
\author{Calum Maitland}
\email{cm350@hw.ac.uk}
\affiliation{Institute of Photonics and Quantum Sciences, Heriot-Watt University, Edinburgh EH14 4AS, UK} 
\affiliation{School of Physics and Astronomy, University of Glasgow, Glasgow G12 8QQ, United Kingdom}
\author{Daniele Faccio}
\email{Daniele.Faccio@glasgow.ac.uk}
\affiliation{School of Physics and Astronomy, University of Glasgow, Glasgow G12 8QQ, United Kingdom}
\author{Francesco Marino}
\affiliation{Consiglio Nazionale delle Ricerche - INO, Via G. Sansone, I-50019 Sesto Fiorentino, Italy}

\begin{abstract}

We investigate the scattering process of Bogoliubov excitations on a rotating photon-fluid.
Using the language of Noether currents we demonstrate the occurrence of a resonant amplification phenomenon,
which reduces to the standard superradiance in the hydrodynamic limit. We make use of a time-domain formulation where superradiance emerges as a transient effect encoded in the amplitudes and phases of propagating localised wavepackets. Our findings generalize previous studies in quantum fluids to the case of a non-negligible quantum pressure and can be readily applied also to other physical systems, in particular atomic Bose-Einstein condensates. Finally we discuss ongoing experiments to observe superradiance in photon fluids, and how our time domain analysis can be used to characterise superradiant scattering in non-ideal experimental conditions.

\end{abstract}

\date{\today}
\maketitle

\section{Introduction}

Superradiant scattering is an effect whereby waves reflected from a moving medium are amplified, extracting energy and momentum in the process. For an axially symmetric and rotating medium such amplification occurs whenever the following superradiant condition is met by the incident wave
\be
\omega < m\Omega \label{E:condition}, 
\ee
where $\omega$ is the wave's angular frequency, $m$ its angular momentum and $\Omega$ is the magnitude of the angular velocity of the rotating object. This phenomenon was famously introduced by Zel'dovich \cite{Zeldovich} in 1971 who showed that low frequency electromagnetic waves reflected by a rotating conducting cylinder are amplified in this manner, whence the terminology `Zel'dovich effect' for this kind of scattering. Following the original Zel'dovich proposal, it was suggested by Misner \cite{misner1972stability}, motivated by previous works by Penrose \cite{Penrose1971}, that this anomalous reflection also occurs near rotating black holes, coining the terminology `superradiance' for this effect.  This led to significant interest in the phenomenon and its deep relation to black hole mechanics \cite{starobinskii1973amplification, starobinskil1973amplification2}.

Superradiant scattering is not specific to the Zel'dovich effect for conductors or its cousin for black holes.  In fluid mechanics this anomalous reflection effect is known as `over-reflection' and has been the subject of a long line of inquiry dating back to the 1950's \cite{Ribner1957,Miles1957,acheson_1976,takehiro_hayashi_1992}, appearing even before the publication of Zel'dovich's work\footnote{For example, equation [4.1] in Ref.~\cite{Miles1957} and the condition $Z<0$ in equation [14] in \cite{Ribner1957}, both derived in 1957 and describing the conditions under which over-reflection occurs are both precisely equivalent the condition \eqref{E:condition} as can be easily checked.
} and culminating recently in a direct experimental observation in a rotating water experiment \cite{overR_prop, overR_exp}.

The parallel between these two communities working on essentially the same physics has remained unknown and is only now coming to light, through developments at a modern interface of gravitation and fluid mechanics known as analogue gravity.  Analogue gravity is the observation that gravitational effects may be simulated in fluid mechanical systems\footnote{Amongst other physical systems: See \citep{LivingReview} and references therein for a modern overview of this interdisciplinary research field.} including quantum fluids, in the lab. In particular, attention has focussed on the simulation of wave scattering effects in the strong gravitational fields that exist near astrophysical black holes \cite{Richartz2009, Silke}.  This possibility has recently been realised experimentally in a draining fluid vortex \cite{Torres}. 


In completely unrelated developments, a new approach to quantum fluids has emerged in the form of  quantum  fluids  of  light,  where effective photon-photon interactions of a monochromatic laser  beam  propagating  in  a  nonlinear  medium  lead  to
a collective behaviour of the many photon system, leading to a superfluid behaviour \cite{Frisch,Chiao1999,Vocke2015,Vocke2015a}.  Such photon fluids have been shown to be suitable candidates for the simulation of analogue curved spacetimes  \cite{Marino2008,Elazar2012}, having received recent experimental attention \cite{Vocke:18}. In particular it has been established that the Kerr metric of a rotating black hole can be realised in a photon fluid, which exhibits a superradiant scattering spectrum \cite{Marino2009}. However, this spectrum applies in the acoustic limit in which the quantum pressure (optical diffraction) is completely neglected and the reflection coefficient decribes the amplitude of continuous plane waves as measured by an observer infinitely far away from the black hole. Neither of these conditions can be truly realised in an experiment.

In the present work we show that superradiant scattering persists in non-linear photon-fluids when the quantum pressure is not negligible; that this effect occurs across the full non-linear dispersion relation and is not specific to the acoustic part of the spectrum where the analogy with gravity is present. We do so through a transient formalism making use of time-dependent wave-packets possessing finite spatial extent, essential for practical analysis of an analogue experiment. This work can be viewed as a natural continuation of work on analogue superradiant scattering in fluids \cite{Bekenstein1998, Cardoso2017,Endlich2017,Visser2005,Richartz2013,Basak2003,Marino2008, Marino2009,berti, lepe,ednilton3}. However, it can also be viewed from the perspective of non-linear optics alone, as a novel phenomenon not previously considered, or as a natural extension of the previous work on over-reflection in ordinary fluids to the case of superfluids such as BEC or photon fluids. We anticipate a close relationship between this effect and other amplification phenomena associated with rotation and angular momentum optics (see for example  Refs.~\cite{faccio1, faccio2,gooding}).

In what follows we will explicitly work with the full optical equations, demonstrating the condition \eqref{E:condition} for optical excitations across the full spectrum, without recourse to an analogy with curved spacetime and show how, in the acoustic \textit{i.e.} superfluid limit, we recover the traditional language and results of superradiance in an analogue spacetime.  Due to a very close relationship between the the governing equations for photon fluids and Bose-Einstein condensates (BEC), the present work also serves as a roadmap for future experiments in BEC seeking to observe superradiant scattering.

\section{Photon fluids}

In this work our photon fluid involves the propagation of a monochromatic optical beam of wavenumber $k$ in a self-defocusing non-linear medium. In the paraxial approximation the slowly varying envelope is described by the well-known non-linear Schr\"{o}dinger equation  (NLSE),
\be
\i\partial_zE+\frac{1}{2k}\nabla^2 E-k\frac{n_2}{n_0}|E|^2 E=0 \label{E:NLSE},
\ee
where $z$ is the direction of propagation, $\nabla^2$ is the transverse Laplacian, $n_0$ is the refractive index and $n_2$ the non-linearity \cite{Marino}. The photon fluid is the (two-dimensional) transverse part of the electric field $E$ while the $z$-direction plays the role of a time coordinate. The laboratory time variable plays no role in this description. To this end we re-write \eqref{E:NLSE} as 
\be
\i\partial_t E+\alpha\nabla^2 E-\beta|E|^2E=0, \label{E:NLSEt}
\ee
where we have introduced the time-like coordinate $t=n_0z/c$ with $c$ the speed of light in vacuum and we have defined the constants
\be
 \alpha=\frac{c}{2n_0k} \quad \text{and}  \quad \beta=\frac{ckn_2}{n_0^2}, 
 \ee
which parametrise the strength of optical diffraction and nonlinearity respectively. In the defocusing case we have case $\beta>0$, while $\alpha>0$ always. The connection with fluid dynamics is drawn when when we express the field $E$ in terms of its amplitude and phase $E=\sqrt{\rho}\,\text{e}^{\i \phi}$. Then \eqref{E:NLSEt} is equivalent to the coupled set
\begin{align}
0&=\dot{\rho}+\nabla\cdot\left(\rho \mbf{v}\right) \label{E:euler1}\\
0&=\dot{\psi}+\frac{1}{2}\mbf{v}^2+c_s^2-\alpha^2\frac{\nabla^2\sqrt{\rho}}{\sqrt{\rho}}. \label{E:euler2}
\end{align}
Apart from the optical coefficients, Eqs.~\eqref{E:euler1} and \eqref{E:euler2} are identical to those describing the density and the phase dynamics of a two-dimensional BEC in the presence of repulsive atomic interactions \cite{stringariBEC}. In this mapping onto superfluids the optical intensity $\rho$ is equivalent to a fluid density, $\mbf{v}:=2\alpha\nabla\phi$ to a fluid velocity and $c_s^2:=2\alpha\beta \rho=c^2n_2\rho/n_0^3$ is the squared sound speed.  The optical nonlinearity, corresponding to the atomic interaction, provides the bulk pressure $P=c^2n_2\rho^2/(2n_0^3)$.\footnote{Note that the sound speed is $c_s^2=\partial P/\partial\rho$, as required.} The last term in \eqref{E:euler2} $\propto \alpha^2$, the quantum pressure, has no equivalent in classical fluid mechanics and arises from the wave nature of light and is due to diffraction. This term is significant in rapidly varying and/or low-intensity regions such as dark-soliton cores and close to boundaries. This set of equations is equivalent to a Klein-Gordon equation in curved spacetime when one neglects this term, as is common in the analogue gravity literature. In this work we retain it for two reasons. Firstly, we will show that superradiant scattering persists beyond the analogy with  gravity and curved spacetime deep into the dispersive regime. Secondly it allows closer accuracy with realistic experiments, in which diffraction can never be entirely ignored.
 
Small perturbations $\psi$ on top of the background beam $E_0$ of the form 
\be
E_\text{total}=E_0\left(1+\psi\right) \label{E:psi_def},
\ee
with $|\psi|\ll|E_0|$, satisfy the Bogoliubov de-Gennes (BdG) equation
\begin{align}
\left(\partial_t+\mbf{v}\cdot \nabla-\i\frac{\alpha}{\rho}\nabla\cdot \rho\nabla\right)\psi+\i\beta\rho\left(\psi+\overline{\psi}\right)=0\label{E:BdG}
\end{align}
when both $E_0$ and $E_\text{total}$ obey the NLSE \eqref{E:NLSEt} (the overline notation $\overline{\psi}$ denotes the complex conjugate of $\psi$).
Note that this choice of perturbation variable $\psi$ differs from the alternative decomposition common in optics $E_\text{total}=E_0+\tilde{\psi}$ by a multiplicative factor of the background field $E_0$. That is, $\psi$ is a dimensionless perturbation while $\tilde{\psi} \equiv E_0 \psi$ is an electric field. 
 
For stationary\footnote{Note that in the optical context stationary means invariant along the propagation direction  $z$, which acts as effective time $t = n_0 z /c$ for the photon fluid.} beams, precisely as in a BEC, the solutions of \eqref{E:BdG} always come in pairs of modes with opposite frequencies,
\be
\psi=\psi_S+\psi_I=U(\mbf{x})\text{e}^{-\i\omega t}+\overline{V}(\mbf{x})\text{e}^{\i\omega t}, \label{E:basicsigidl}
\ee
which we label as positive $\propto e^{-i \omega t}$ and negative  $\propto e^{i \omega t}$ modes. In regions of constant flow and intensity ($\mbf{v}=$const, $\rho=$const) or equivalently in a WKB approximation\footnote{This approximation is valid whenever the wavelength of a mode is much shorter than the typical length scale on which the background flow and intensity is varying and provides the intuitive picture of a wave which slowly changes its amplitude and phase on an inhomogeneous background.}, local plane wave solutions with $U=\text{e}{}^{\i\mbf{k}\cdot\mbf{x}}$ to the BdG equation must satisfy the familiar Bogoliubov dispersion relation
\be
\left(\omega-\mbf{k}\cdot \mbf{v}\right)^2-\alpha^2 \mbf{k}^4-c_s^2\mbf{k}^2=0.  \label{E:bogo}
\ee
Note that this is a position dependent dispersion relation, depending on the location in the flow through $\mbf{v}(\mbf{x})$ and $c_s(\mbf{x})$. 
Including the frequencies, the complete positive mode 
\be
\psi_S=\text{e}^{\i \mbf{k}\cdot{\mbf{x}}}\text{e}^{-\i\omega t}
\ee
labelled by the pair $(\omega, \mbf{k})$ can be shown to locally induce a negative mode labelled by $(-\omega, -\mbf{k})$
\be
\psi_I=\zeta\, \text{e}^{-\i \mbf{k}\cdot{\mbf{x}}}\text{e}^{\i\omega t} \label{E:modes}.
\ee
The constant of proportionality $\zeta$ can be explicitly calculated to be 
\be
\zeta=\frac{1}{\beta\rho}\left(\omega-\mbf{k}\cdot\mbf{v}-\alpha\mbf{k}^2-\beta\rho\right) \label{E:zeta}
\ee
as can be easily verified by inserting this pair into \eqref{E:basicsigidl} then eventually \eqref{E:BdG}, neglecting derivatives of the background quantities.

\begin{figure}
\centering
\includegraphics[width=\linewidth]{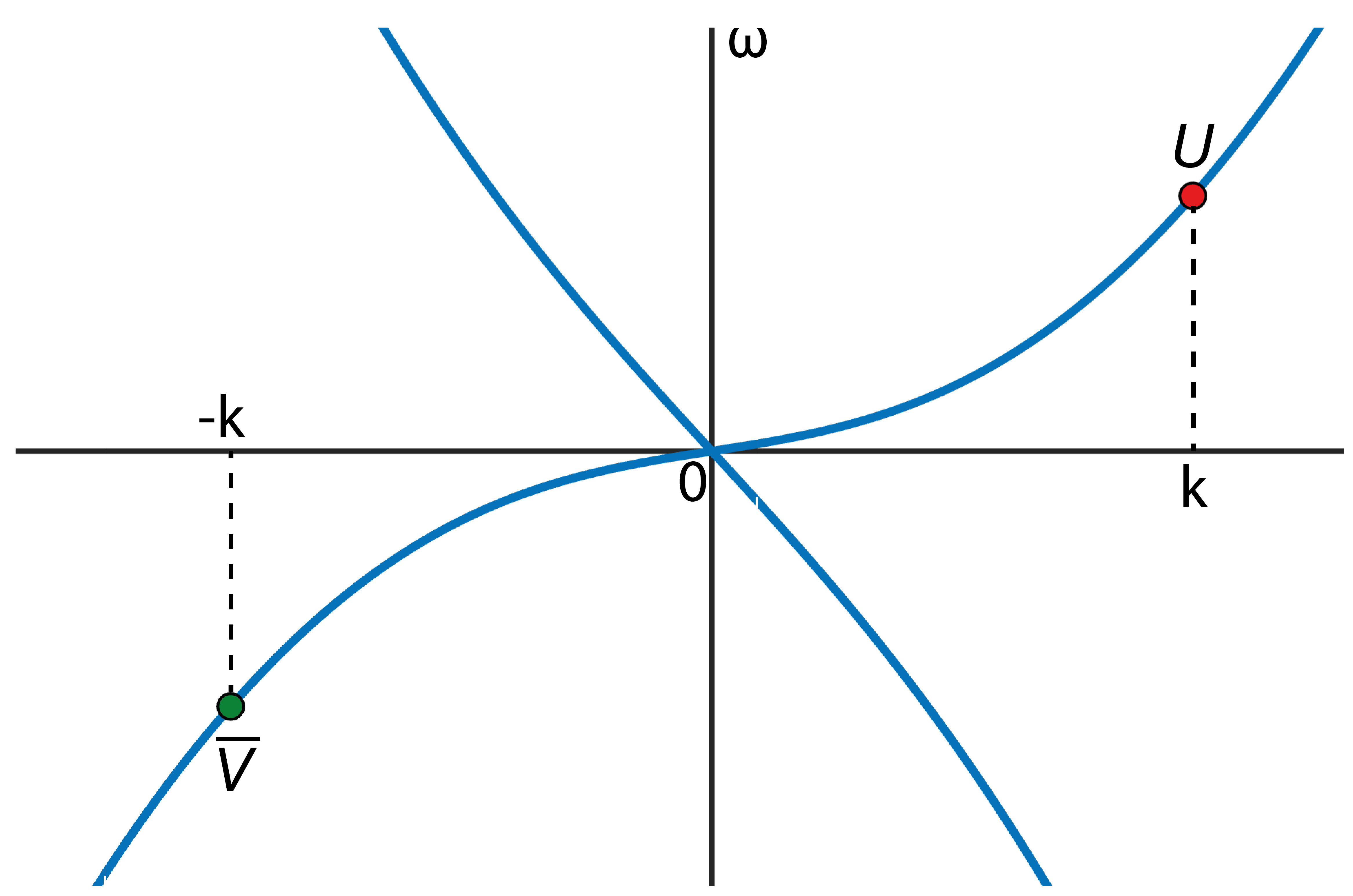}
\caption{Bogoliubov dispersion curve for an acoustic probe of the photon fluid, including a component of the fluid flow along the probe's propagation direction ($\mbf{v}=(-v,0)$). 
The positive $U$ and negative $\overline{V}$ modes are indicated with red and green circles respectively, which together form coupled pairs of solutions to the BdG equation. The coupling between these modes is provided by the optical nonlinearity.
 \label{F:dispersion}}
\end{figure}

The relation \eqref{E:bogo} is symmetric under the simultaneous replacement $\mbf{k}\rightarrow -\mbf{k}$ and $\omega\rightarrow-\omega$ and therefore negative modes also satisfy the Bogoliubov dispersion relation, while sitting in a different location on the dispersion curves as we show in Fig.~[\ref{F:dispersion}]. 
Note that both positive and negative modes always have the same group and phase velocity and hence always travel in the same direction with the same speed.

These observations will play an important role in the calculations in the following section so are worth re-emphasising. Normal mode solutions to the BdG equation always come in positive-negative pairs. These pairs are locally `almost' complex conjugates of one another: they are of opposite phase but there is an amplitude difference given locally by $\zeta$ in \eqref{E:zeta}. 

We finish this section with a short note to clarify the connection with real optical experiments.  Only measurements of the total field $E_\text{total}$ can be made in real experiments, which is the sum of two components $E_0$ and $E_0\psi$.  By conducting an unperturbed  $\psi=0$ experiment, $E_0$ is known and can be subtracted from $E_\text{total}$.  What is left in the measured electric field is the product $E_0\psi$, which from \eqref{E:psi_def} and \eqref{E:basicsigidl} we have seen will consist of the pair
\begin{align}
E_S&=E_0\psi_S \label{E:optsignal}\\
E_I&=E_0\psi_I. \label{E:optidler}
\end{align}
%
What typically occurs in optical experiments is that the background field $E_0$ (the `pump' beam) is provided by one laser with a second independent laser providing the electric field perturbations $E_S$, $E_I$. In this way the variable $\psi$ which satisfies the Bogoliubov dispersion is a mixture of the pump and probe beams with a phase which is the sum of the pump and probe phase and intensity which is the product.  For this reason the measured fields $E_S$ and $E_I$ do not satisfy the same Bogoliubov dispersion relation.

\section{Superradiance in photon fluids} \label{S:super}

We will now demonstrate that the BdG equation \eqref{E:BdG} possesses superradiant scattering solutions in general (for all values of $\alpha$) and not just in the acoustic limit.  We will do this by monitoring the radial propagation of a localised annular wavepacket, comprised of the individual positive and negative wavepackets, as it approaches and scatters from an optical Zel'dovich cylinder. We will make use of a conserved quantity (the conservation relation will be proved in a later section) to relate the reflected and transmitted wavepackets.

Consider the idealised background (pump) beam profile $\rho=$const and
\be
v_\theta=\begin{cases} r\Omega, & r<r_0 \\ 0,& r>r_0 \end{cases} \label{E:Zel}, \qquad v_r=0,
\ee
with $\Omega=$constant. Such a beam is the analogue of the rigidly rotating circular body on which the traditional Zel'dovich effect occurs. We choose this background as a simple example for which the analysis is straightforward. It is also an opportunity to introduce methods which later we shall use in more general backgrounds after having established the existence of the basic superradiant mechanism here.   

\begin{figure}
\centering
\includegraphics[width=\linewidth]{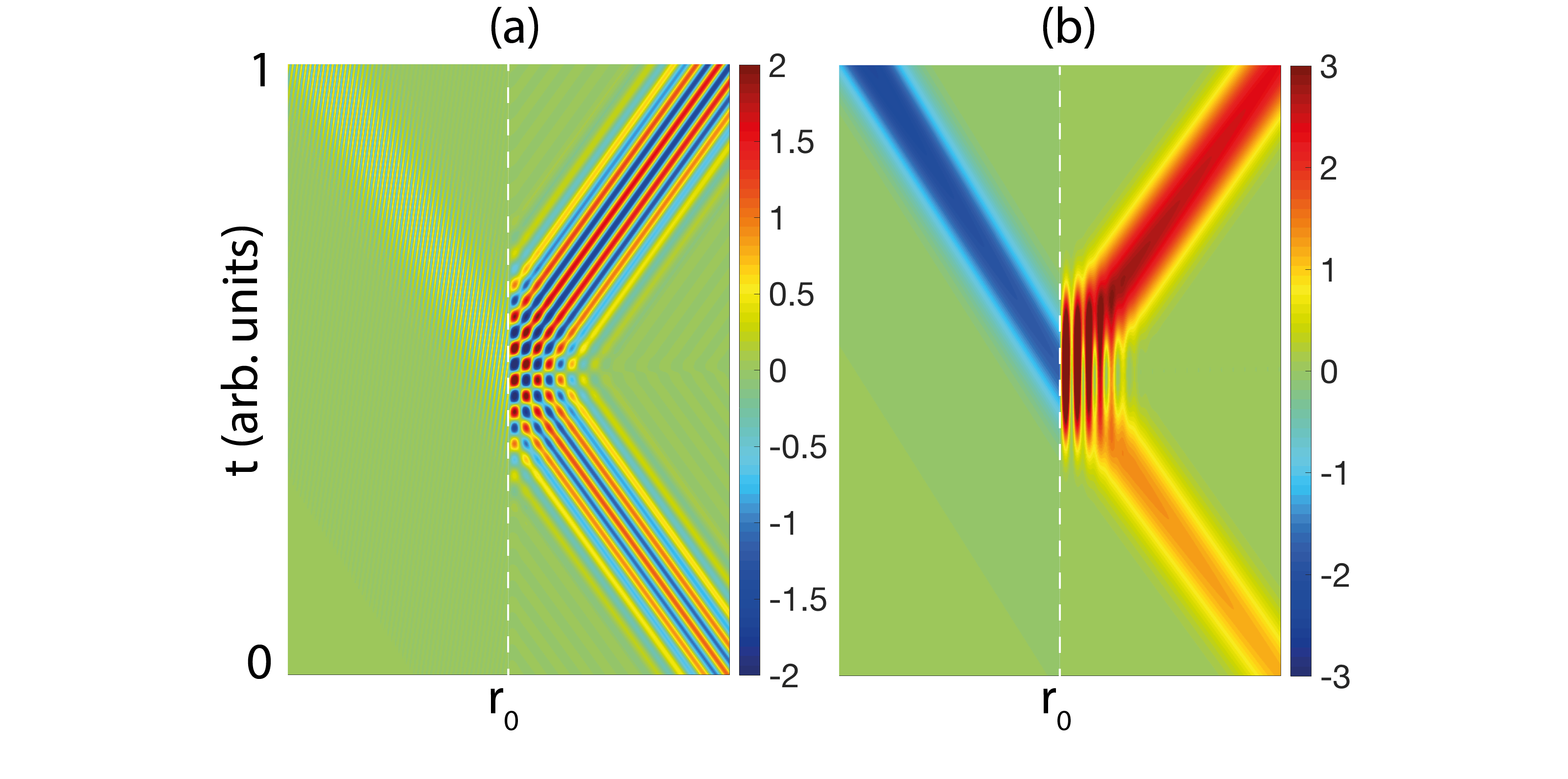}
\caption{Real part of the positive mode $\Re \left(U(r,t)\right)$ under going superradiance (subfigure a) and associated current/Noether charge density $J_0(r,t)$ as defined in equation \eqref{E:dcurr} (subfigure b). The white dashed line indicates the edge of the Zel'dovich cylinder $r=r_0$. Note that the transmitted modes have the peculiar property that their phase velocity points in the opposite direction to the group velocity as can be seen in subfigure a. 
 \label{F:2}}
\end{figure}

Consider now, on top of this background, an annular positive mode wavepacket $\psi_S$ of fixed positive frequency $\omega>0$ and positive angular momentum $m>0$, initially in the region $r>r_0$, propagating inwards towards the rotating `body' as in the lower portion of Fig.~[\ref{F:2}]. Associated with this positive mode is of course a negative mode $\psi_I$ which completes the exact solution to \eqref{E:BdG} 
\be
\psi = \psi_S + \psi_I.
\ee
Due to the angular symmetry and stationarity of the background, the frequencies and angular momenta of the positive and negative modes are conserved and of opposite sign
\begin{align} 
\psi_S &= U(r)\text{e}^{-\i\omega t+\i m \theta} \notag \\
\psi_I &= \overline{V}(r) \text{e}^{+\i\omega t-\i m \theta} \label{E: sigidlsplit}.
\end{align}
The functions $U$ and $\overline{V}$ implicitly carry the Gaussian envelope and radial momentum of the packets. 

\begin{figure}
\centering
\includegraphics[width=\linewidth]{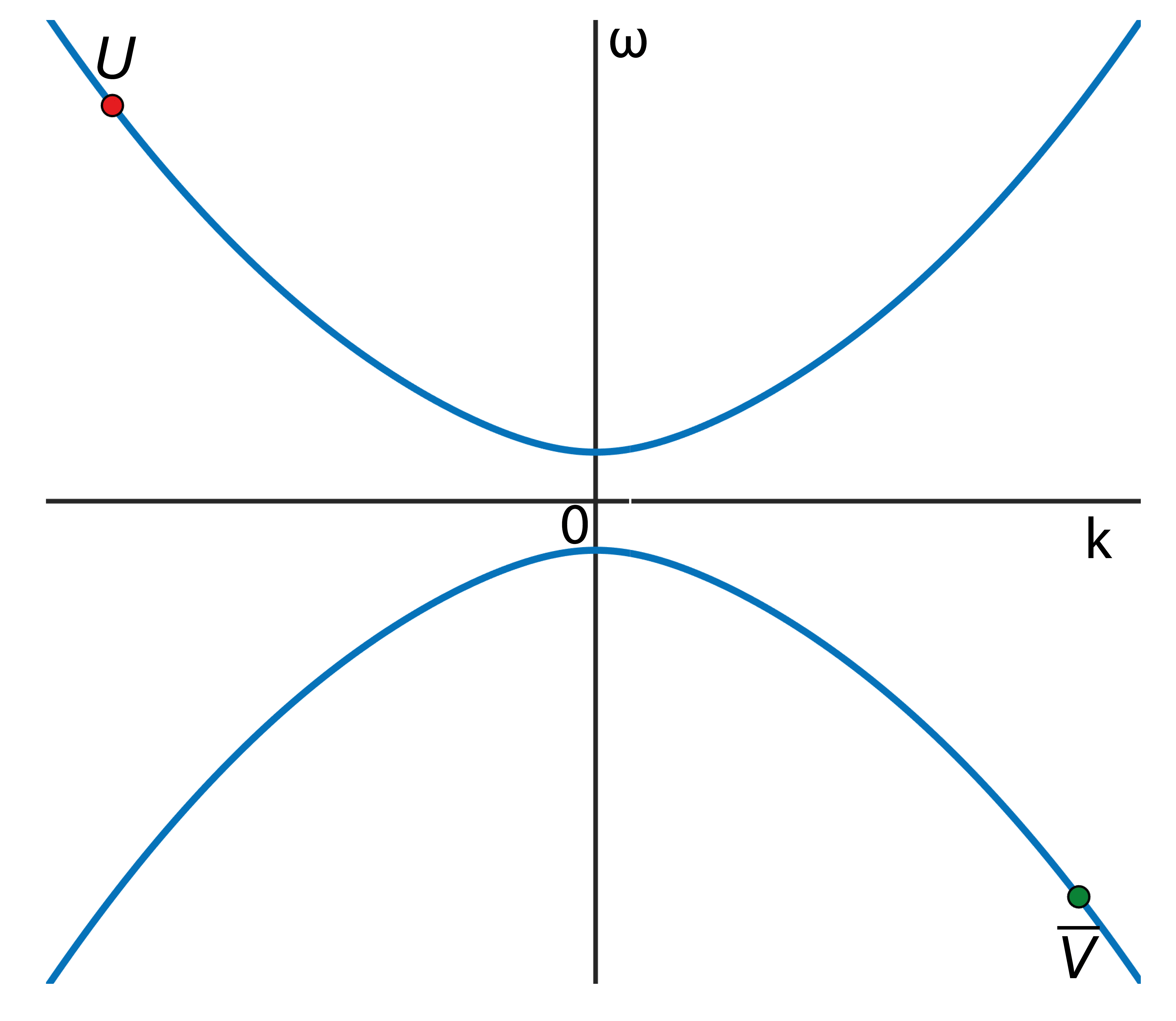}
\caption{The initial positive $U$ (red circle) and negative $\overline{V}$ (green circle) modes' locations on the BdG dispersion in the exterior region, with $U$ and $\overline{V}$ as specified by equations \eqref{E:earlyU} and \eqref{E:earlyV} respectively. Both modes have negative group velocity and hence move radially inwards towards the cylinder. \label{F:exterior}}
\end{figure}

We can without loss of generality assume that locally these initial wavepackets, being in the region of zero flow, form parts of propagating Bessel modes i.e. Hankel functions \footnote{It is straightforward to show that the inward propagating Hankel function pair 
\be
f=H^+_m(kr)\text{ e}^{-\i\omega t+\i m \theta}+\zeta	 H^-_m(kr)\text{ e}^{+\i\omega t-\i m\theta}.
\ee
exactly solves the BdG equation in the exterior zero flow region with $\mbf{k}$ satisfying \eqref{E:bogo} and $\zeta$ given by \eqref{E:zeta}.}, and that within the initial wavepacket envelope we may approximate
\be
U(r)\simeq \mbf{1}\times\frac{\text{e}^{-\i k r}}{\sqrt{r}}. \label{E:earlyU}
\ee
The wavenumbers $\omega,k, m$ are not independent, being determined by the dispersion relation \eqref{E:bogo} in the $r>r_0$ region which, for our Zel'dovich background, is shown in Fig.~[\ref{F:exterior}]
where we note the `mass gap' arising from the angular momentum $m>0$. 
Using the same arguments that led to \eqref{E:zeta}, initially we have that
\be
\overline{V}\simeq \mbf{1}\times \frac{1}{\beta\rho}\left(\omega-\alpha\tilde{k}^2-\beta\rho\right)\frac{\text{e}^{+\i k r}}{\sqrt{r}} \label{E:earlyV}
\ee
where $\tilde{k}^2=k^2+(m^2-1/4)/r^2$ is the local wavenumber\footnote{This locally varying wavenumber arises from the approximation of a Bessel function by an exponential. The full Bessel function has the constant Laplacian eigenvalue $-k^2$ for the radial wavenumber.} and the coefficients $\mbf{1}$ carry implicitly the localised Gaussian envelopes. Recall that the positive and negative modes have the same phase and group velocities and hence both move towards the rotating flow. 

With this description of the initial condition we now let the wavepackets propagate and scatter from the rotating cylinder. At late times, the incident wavepacket has split into reflected and transmitted components which are separated in space. Due to the stationarity and axisymmetry, the quantum numbers $m$ and $\omega$ are conserved throughout the scattering so that at late times the solution is again of the form \eqref{E: sigidlsplit} but now with the two oppositely propagating wavepacket components in both the positive and negative channels:
\begin{align}
U(r)&=\mbf{R}_U(r)+\mbf{T}_U(r) \\
\overline{V}(r)&=\overline{\mbf{R}}_V(r)+\overline{\mbf{T}}_V(r) .
\end{align}
The exterior reflected components again take the known Bessel form which we again approximate with exponentials
\begin{align}
U(r)&\simeq R_U\,\frac{\text{e}^{+\i k r}}{\sqrt{r}}+T_U(r) \label{E:U} \\
\overline{V}(r)&\simeq\overline{R}_V \,\frac{\text{e}^{-\i k r}}{\sqrt{r}}+\overline{T}_V(r) .
\end{align}
In Fig.~[\ref{F:2}] we show a representative simulation of \eqref{E:BdG} plotting only the positive wavepacket $U$ as it propagates and scatters from the edge of the cylinder at radius $r_0$.


Using the local proportionality result \eqref{E:zeta} (i.e. that positive and negative modes are locally conjugates of each other, up to a real constant) we know that the reflected components of the positive and negative modes satisfy
\be
\overline{R}_V=\overline{R}_U\times \frac{1}{\beta\rho}\left(\omega-\alpha\tilde{k}^2-\beta\rho\right)
\ee
so that we have
\begin{align}
\overline{V}(r)\simeq \overline{R}_U\,\frac{1}{\beta\rho}\left(\omega-\alpha\tilde{k}^2-\beta\rho\right)\frac{\text{e}^{-\i k r}}{\sqrt{r}}+\overline{T}_V(r) .
\end{align}
We have not explicitly written down the radial phases of the transmitted wavepackets $T_U$ and $\overline{T}_V$ as we were not able to find simple expressions - they are the normal modes of the rotating disc background which are or Bessel-like form. Their exact forms are unimportant however, since in order for the pair of transmitted wavepackets to locally be a solution to \eqref{E:BdG} the relationship between the transmitted modes must be
\be
\overline{T}_V=\frac{1}{\beta\rho}\left(\omega-m\Omega+\alpha\nabla_m^2-\beta\rho\right)\overline{T}_U,
\ee
where $\nabla^2_m=r^{-1}\partial_r r \partial_r-m^2/r^2$. Assuming that our transmitted wavepacket is sufficiently localised we can, without loss of generality, assume that $\overline{T}_U$ is locally an eigenfunction of $\nabla_m^2$ with a (possibly $m$ dependent) eigenvalue\footnote{There is the possibility of complex eigenvalues here which will occur if the transmitted modes are too close to the origin which can be avoided by choosing $r_0$ sufficiently large.} $-\lambda^2_m$. We therefore have
\be
\overline{T}_V=\frac{1}{\beta\rho}\left(\omega-m\Omega-\alpha\lambda_m^2-\beta\rho\right)\overline{T}_U.
\ee
Hence the full solution at late times is given by $\psi_S+\psi_I$ decomposed as in \eqref{E: sigidlsplit} with $U(r)$ as in \eqref{E:U} and 
\begin{align}
\overline{V}(r)&= \overline{R}_U\times \frac{1}{\beta\rho}\left(\omega-\alpha\tilde{k}^2-\beta\rho\right)\frac{\text{e}^{-\i k r}}{\sqrt{r}}\notag \\
&\hspace{10mm}+\frac{1}{\beta\rho}\left(\omega-m\Omega-\alpha\lambda_m^2-\beta\rho\right)\overline{T}_U(r) \label{E:late}
\end{align}

We wish to relate the power of the reflected positive mode $R_U$ to the power of the incident positive mode, which we have chosen here to be 1. In particular we want to find conditions under which the reflected power can be greater than the incident power, so that superradiance can occur.  To this end we need to relate the solutions at early times to the solution at late times. This will be achieved by making use of a conservation relation, which provides a quantity which is conserved over the duration of the evolution.  By evaluating this quantity at early and late times and equating the two we can explicitly relate the reflected and incident amplitudes.  

We will prove the following conservation relation in a subsequent section:
\be
\int r\,dr \left(|U|^2-|V|^2\right)=\text{const}. \label{E:conserv}
\ee
Assuming that this relation holds we proceed by evaluating the left hand side at early  and late times by inserting the asymptotic forms (early \eqref{E:earlyU}, \eqref{E:earlyV} and late \eqref{E:U}, \eqref{E:late}). We find after some algebra the relation
\be
1=|\mathcal{R}|^2+\sigma^2\frac{\left(\omega-m\Omega \right)}{\omega}|\mathcal{T}|^2\label{E:super}
\ee
where we have introduced the total integrated powers $|\mathcal{R}|^2=\int r\,dr |R_U|^2$ (and similarly for $\mathcal{T}$) and
\be
\sigma^2=\frac{\alpha\lambda_m^2+\beta\rho}{\alpha\tilde{k}^2+\beta\rho}
\ee
is a positive constant.

 
This final relation \eqref{E:super} is our main result and shows that $|\mathcal{R}|>1$ whenever the superradiance condition \eqref{E:condition} holds, as claimed at the beginning of this section. 

\begin{figure}[h] 
\centering
\includegraphics[width=0.8\linewidth]{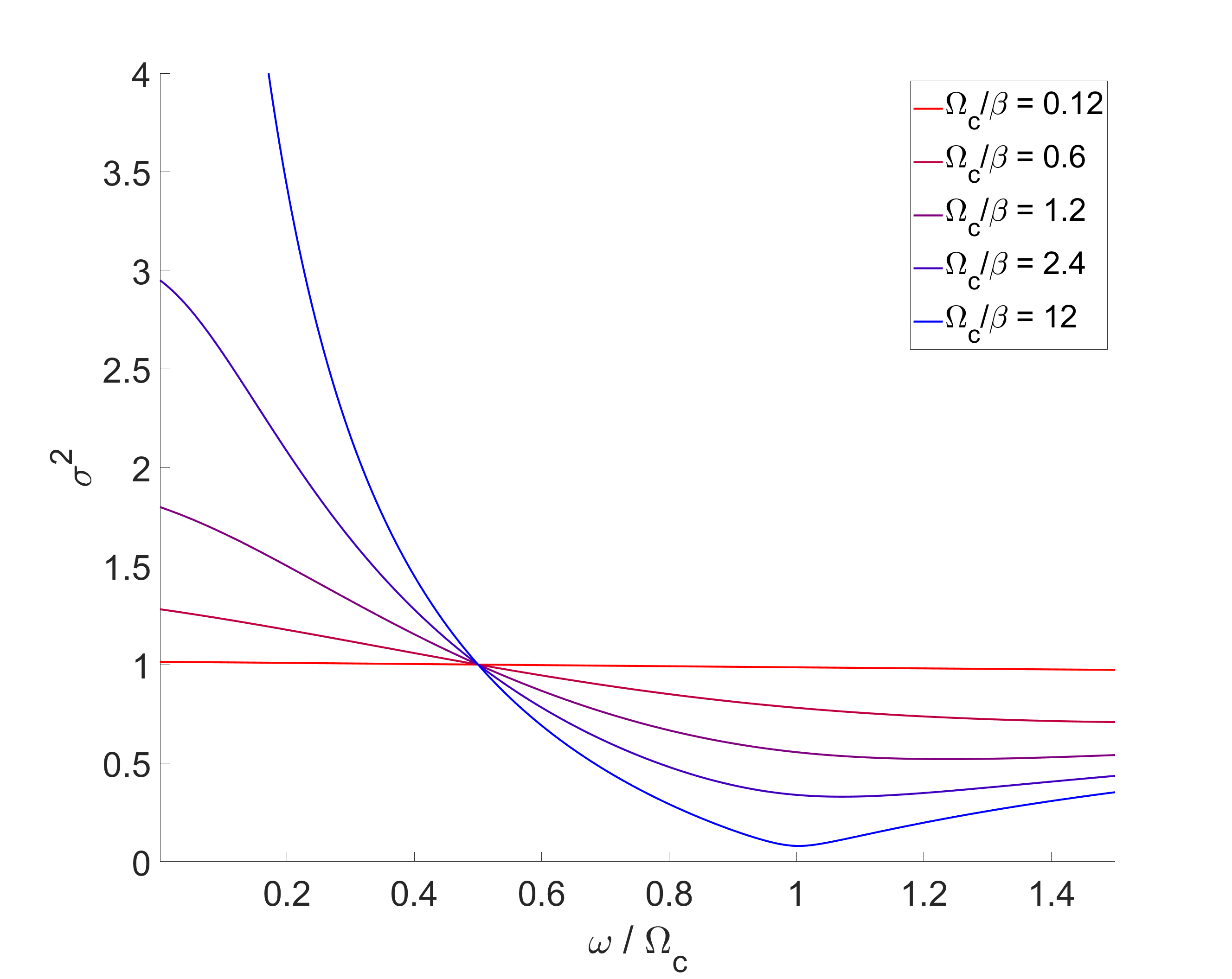}
\caption{\small{Modification $\sigma^2$ of the overreflection coefficient vs. frequency, scaled to the overreflection cut-off frequency $\Omega_c = m \Omega$, for different values of $\Omega_c / \beta \rho$.} }
\label{F: sigmaplot}
\end{figure}

We note that this result is completely general, holds even if we are far from the `acoustic' or linear part of the dispersion relation \eqref{E:bogo} and does not require any radial flow structures.  The result generalises previous work in this aspect. Fig.~[\ref{F: sigmaplot}] plots $\sigma^2$, which characterises the deviation from classical superradiant reflection, for varying the strength of $\beta$ (proportional to the optical nonlinearity supporting the photon fluid). As $\beta \rightarrow \infty$, $\sigma^2 \rightarrow 1$ for all phonon frequencies, recovering the acoustic limit.

\begin{figure}
\centering
\includegraphics[width=\columnwidth]{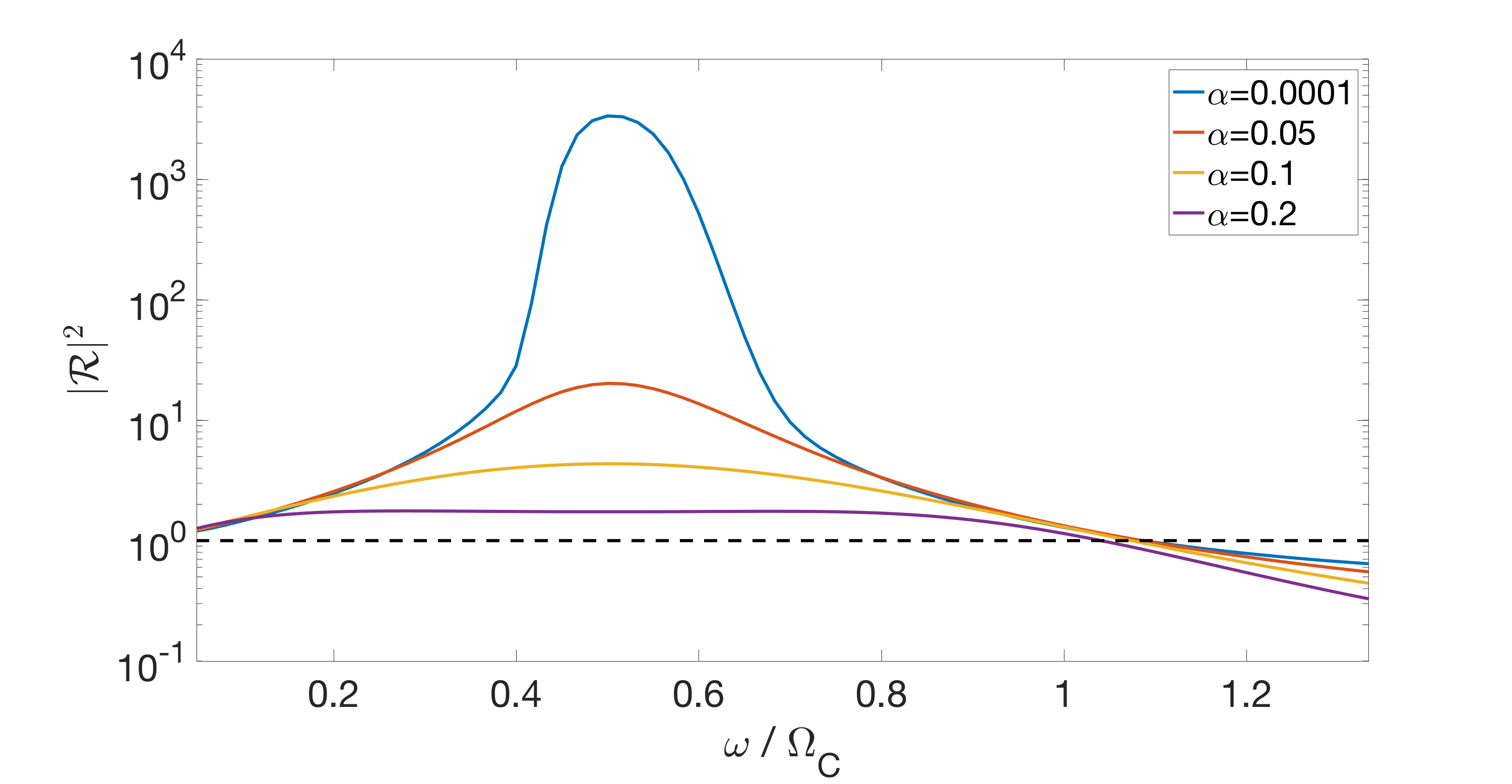}
\caption{Superradiant scattering of Gaussian wavepackets from an optical Zel'dovich cylinder. The reflected amplitude is measured using our integrated current method developed in Sec.~\ref{S:trans}.  
\label{F:spectrum}}
\end{figure}

In Fig.~[\ref{F:spectrum}] we show the results of numerical simulations of the BdG equation \eqref{E:BdG} by plotting the spectrum of $|\mathcal{R}|^2$ as a function of the initial $\omega$ for a range of values of the quantum pressure parameter $\alpha$.  Note that $\alpha\rightarrow 0$ with $\alpha\beta=$const being the acoustic limit. 
We see the presence of the critical cut-off frequency above which superradiance ceases, independently of the quantum pressure parameter $\alpha$ and also that superradiance appears to be more dramatic as one approaches the acoustic limit $\alpha\ll 1$. There also appears to be a resonant effect which amplifies the reflection at half the cut-off frequency, for which $\sigma^2=1$ as the wave-numbers $\lambda_m$ and $\tilde{k}$ of modes on either side of the cylinder boundary match. This resonance becomes more pronounced as $\alpha$ decreases. We point out that this resonance is well known and corresponds to the classical transmission resonance from quantum mechanics (see for example \cite{Boonserm2011}) for this particular sharp background configuration.


What we have shown is that superradiance is possible in photon fluids and it does not require an acoustic approximation - the superradiance condition for amplification depends only on the initial frequency and momentum of the mode as well as the rotational velocity of the reflecting medium. The result intrinsically involved the coupled positive-negative mode structure of the solutions and a conservation relation which required knowledge and monitoring of both parts of the solution. This is in contrast to many results in the acoustic approximation in BEC and earlier work in photon fluids whereby only the positive component was considered. Interestingly the generalisation to the dispersive case requires one to measure both components. 

We have carried out this demonstration in a particular simple background flow as a proof of principle for superradiance in photon fluids. We believe the mechanisms and structures involved in this version are not specific to this background, the exploration of dispersive superradiance on more general backgrounds is left for future work, but we describe below a generalisation of superradiance which is naturally applied to modern optical experiments.



%


\section{Conserved currents and a general definition of superradiance} \label{S:trans}

In this section we will prove the crucial conservation relation \eqref{E:conserv} using a symmetry argument from an action principle for the BdG equation. In the previous section we used this relation to demonstrate superradiance in a single monochromatic scattering event on a particular background. 

The machinery we introduce here is very powerful and we will show how it can be used to characterise superradiant scattering in a much more general context than this monochromatic, `perfect' wavepacket scattering example including in non-stationary backgrounds. 

We will show that the conservation relation \eqref{E:conserv} arises as the conserved charge associated with a symmetry in an action giving rise to the BdG equation \eqref{E:BdG}. By partially integrating the resulting Noether conservation relation down to a finite radius one obtains a measure of the total Noether charge in the integrated region. This `exterior charge' is not positive definite nor conserved and increases when superradiance occurs. We show that this coincides with a transmitted mode with negative Noether charge escaping the integrated region. Based on this new perspective we propose a more general definition of superradiance which is applicable to modern non-linear optical experiments which reduces to the standard one in the traditional setting.




For simplicity we again work in the special Zel'dovich case described in \eqref{E:Zel} to outline the idea. Since the background is axially symmetric, the positive-negative mode decomposition of the perturbation takes the general form
\be
\psi=U(t,r)\text{e}^{\i m \theta}+\overline{V}(t,r)\text{e}^{-\i m \theta} \label{E:signal_idler}
\ee
where we have absorbed the frequencies into the $U$ and $\overline{V}$ amplitudes.  Even if the flows given in \eqref{E:Zel} are idealised, such a decomposition will hold in more general backgrounds including under non-stationary and finite-size experimental conditions. 

By some simple manipulations Eq.~\eqref{E:BdG} is equivalent to the following de-coupled pair of higher order partial differential equations for the amplitudes $U$ and $\overline{V}$
\begin{widetext}
\begin{align}
&\left(\partial_t+\i m \Omega+\i\alpha \nabla^2_m\right)\left(\partial_t+\i m \Omega-\i\alpha \nabla^2_m\right)U-2\alpha\beta\nabla^2_mU=0\notag \\
&\left(\partial_t-\i m \Omega+\i\alpha \nabla^2_m\right)\left(\partial_t-\i m \Omega-\i\alpha \nabla^2_m\right)\overline{V}-2\alpha\beta\nabla^2_m\overline{V}=0. \label{E:signal_idler_eqns}
\end{align}
Here we introduced the shorthand $\tilde{\Omega}(r)=\Omega\, \text{Heaviside}(r_0-r)$ and then for simplicity dropped the tildes and the functional dependence on $r$, leaving these implicit.  We shall refer to these as the BdG-wave equations as they are second order in time partial differential equations with fourth order spatial derivatives. These equations follow as the Euler-Lagrange equations for the following action, 
\begin{align}
S:=\int r drdt  \, \left[\left(\partial_t-\i m \Omega-\i \alpha \nabla_m^2\right)\overline{V}  \right]\times \left[\left(\partial_t+\i m \Omega-\i\alpha \nabla_m^2\right) U\right]-2\alpha\beta\left[\left(\partial_r U\right)\left(\partial_r \overline{V}\right)+m^2U\overline{V}\right] \label{E:action}.
\end{align}
Moreover this action \eqref{E:action} is invariant under the global $U(1)$ phase shift
\be
\begin{pmatrix} U\\ V\end{pmatrix}\longrightarrow \begin{pmatrix} U\\ V\end{pmatrix}\text{e}^{\i\lambda}\simeq\begin{pmatrix} U\\ V\end{pmatrix}+\i\lambda\begin{pmatrix} U\\ V\end{pmatrix}. \label{E:symm}
\ee
\end{widetext}
The Noether current associated with this symmetry has components 
which 
can be simplified, using the BdG equations of motion, to 
\begin{align}
J^0&=-\i \beta\left(|U|^2-|V|^2 \right) \notag \\
J^1&=\alpha\beta\left(V\partial_r\overline{V}-\overline{V}\partial_r V-\overline{U}\partial_rU+U\partial_r\overline{U}\right). \label{E:dcurr}
\end{align}
The current $J^\mu$ satisfies the standard conservation relation $\partial_\mu J^\mu=0$ and therefore the integrated total charge $Q:=\int r dr  J^0$ is constant in time. This is precisely the result we set out to prove. 

This establishes \eqref{E:conserv} from an interesting new perspective to the traditional one: \textit{The Bogoliubov normalisation condition \eqref{E:conserv} can be understood as a conserved Noether charge $Q$ associated with a $U(1)$ symmetry \eqref{E:symm} in the action for the decoupled `BdG-wave' equations. }

Given the conserved current, one can integrate it over an annular sub-region with boundary instead of all space.  The resulting relation
\be
\partial_t\left(\int_\text{region} rdr \,J^0 \right)+\left. J^1\right|_{\text{boundary}}(t)=0  \label{E:leak}
\ee
tells us that as long as $J^1$ is zero at the boundary the total charge in the integrated region is conserved. For example, when a wavepacket moves across the boundary the total charge in the region will change, and that change is governed by \eqref{E:leak}.


Crucially, under special circumstances the flux $J^1|_\text{boundary}$ can be negative. This occurs when a mode with negative Noether charge exits the region. These special circumstances reduce to \eqref{E:condition} when such a condition makes sense but is more general than \eqref{E:condition}, applying also to non-stationary multi-mode situations without the need to decompose into angular $m$-eigenmodes.

As an example let us consider the idealised wavepackets of Sec.~\ref{S:super}. Such modes are characterised by triples $(\omega, m, k)$ for the positive modes and $(-\omega, -m, -k)$ for the negative ones. Let us call the total charge of the initial condition $Q_i$ and let us normalise the solution by dividing by $Q_i$ so that the total initial charge is 1. Then the charge of the final solution is of course the sum of the charges of the reflected and transmitted components since the integral splits into disjoint sub integrals
\be
Q_f=Q_f^T+Q_f^R.
\ee
Of course, since the charge is conserved we have $Q_f=Q_i$. By direct calculation one can show that 
\be
Q^T_f=\sigma^2\frac{\left(\omega-m\Omega \right)}{\omega}|\mathcal{T}|^2
\ee
in the notation of \eqref{E:super}. Therefore we see the component structure of the relation \eqref{E:super} emerging. The left hand side is the (normalised) initial charge while the right hand side is the (normalised) final charge, which is the sum of the reflected and transmitted charge.  The sum of the reflected and transmitted charge is always equal to the initial charge and whenever $\omega<m\Omega$ the transmitted charge is negative while the reflected charge is greater than 1.

In Fig.~[\ref{F:2}] we plot the real part of the positive mode and the local charge of the full solution\footnote{The charge of the positive mode alone is not a meaningful quantity and only its combination with the negative mode has a well-defined charge, see \eqref{E:conserv}.} (scaled by the initial charge). We see that the transmitted charge is negative while the reflected charge is greater than the incident charge.   

\begin{figure}
\centering
\includegraphics[width=\linewidth]{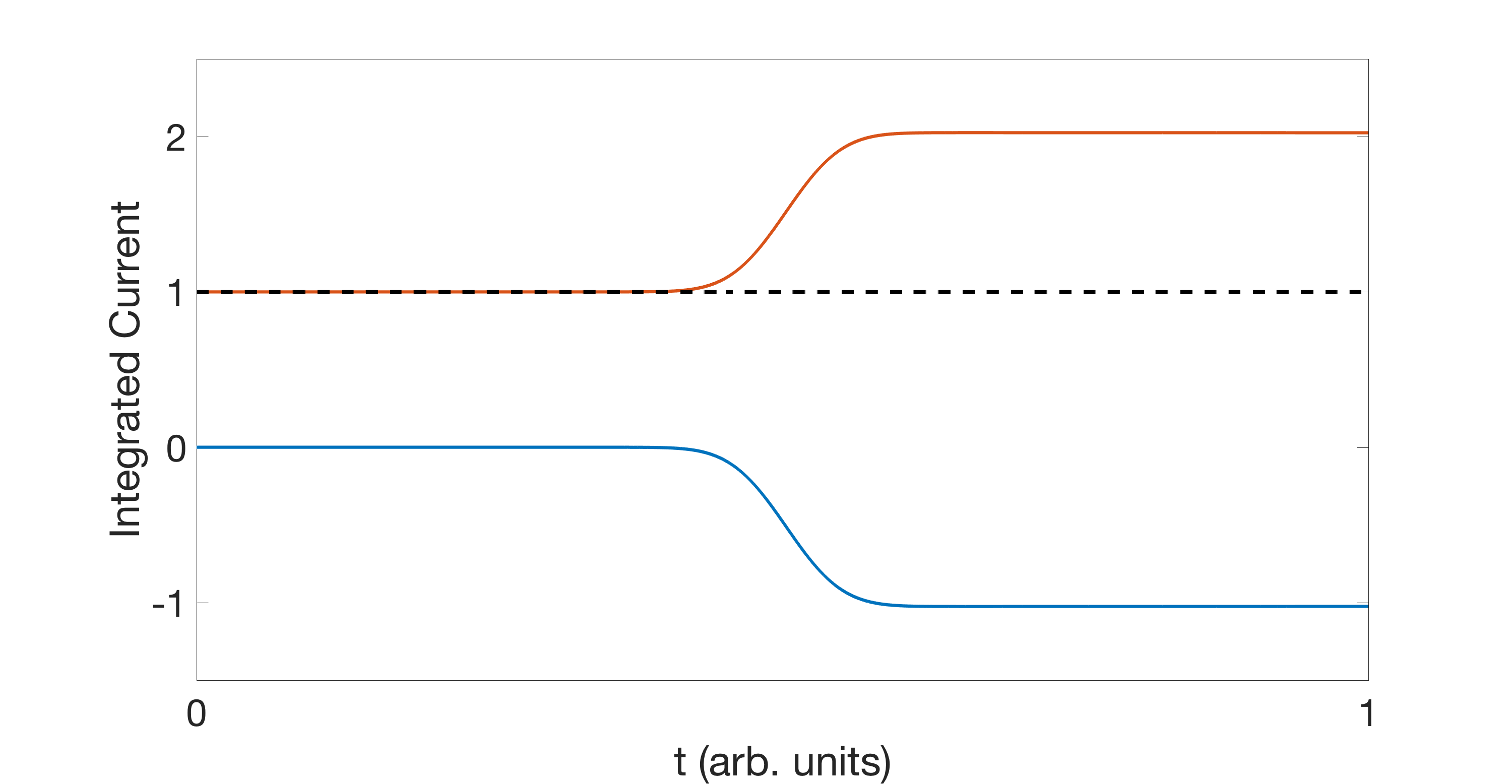}
\caption{Integrals of the Noether charge density shown in Fig. \ref{F:2} \textbf{(b)} inside (blue, lower curve) and outside (red, upper curve) the Zel'dovich cylinder over time, relative to the total charge density at $t=0$. The straight dotted black line shows the global current density integral, showing that total Noether charge is conserved throughout the scattering. } \label{F:integrated}
\end{figure}

Integrating the charge density up to the radius $r_0$ and plotting this over time one observes an increasing function in time. In this idealised example the total exterior charge as a function of time is a very simple step function as shown in Fig.~[\ref{F:integrated}].
In that figure we have also shown the integration of the charge density over the complement region $r<r_0$. In that region there is initially zero charge but this decreases when the negatively charged transmitted modes enter the integration region. 

These observations allow us to make an operational definition of superradiance: superradiance occurs whenever the total charge in an exterior region increases in time. The total charge is measured by \eqref{E:conserv} which can be obtained by measuring only the amplitudes of the positive and negative modes at a single instant. The condition defined by \eqref{E:condition} applies to monochromatic waves in stationary situations. Under those conditions our new definition is precisely equivalent to the standard one \eqref{E:condition}. Our new definition is much more general, allowing for arbitrary initial conditions, incomplete scattering and non-stationary backgrounds. This is perfectly suited to real optical experiments where initial conditions for the perturbations might cover the whole region$r>r_0$ and $r<r_0$ and be complicated waveforms. Looking at Fig.~[\ref{F:integrated}] we also note that it would not be necessary to wait until the full scattering has taken place in order to conclude superradiance. Experimental systems ordinarily have losses which constrain the time scales in which modes can propagate without significant distortion. Freeing ourselves from the confines of the asymptotic analysis "at infinity" by moving to a transient formulation we bring the theory in line with realistic experiments and the possibility of experimental detection of superradiance.


\section{Relationship with analogue models and rotating black holes} \label{S:BH}

As we mentioned in the introduction, there is a well-known analogy between wave propagation in strong gravitational fields and the propagation of excitations in (possibly effectively) moving media. In practice,  and under certain approximations, excitations obey the d'Alembertian equation associated with an effective spacetime metric,
\be
0=\Box_g\psi=\frac{1}{\sqrt{\mbf{g}}}\partial_\mu\left(\sqrt{\mbf{g}}\,g^{\mu\nu}\partial_\nu\psi \right), \label{E:dAl}
\ee
where the effective metric $g_{\mu\nu}$ is a D+1 dimensional matrix with inverse $g^{\mu\nu}$ and determinant $\mbf{g}$. The metric coefficients are functions of the background upon which the excitations propagate (see \cite{LivingReview} for a review). Normally the analogy is valid in an approximation of the perturbation dynamics, one in which the dispersion relation takes a linear `relativistic' form with a universal propagation speed. Indeed the dispersion relation associated with \eqref{E:dAl} is the quadratic $g^{\mu\nu}k_\mu k_\nu=0$, and does not contain the quartic terms of \eqref{E:bogo} for example. 

The strength of our results in the previous sections lies precisely in their validity beyond the acoustic approximation, which we no longer need to monitor throughout a scattering in order to trust the result. Our superradiance result for wavepacket propagation \eqref{E:super} holds over the entire Bogoliubov spectrum, the price we pay for the generality is that both positive and negative modes need to be measured and included in the analysis. As a by-product of this formulation we were able to talk about characterisation of generalised superradiance for arbitrary perturbations. The purpose of this section is to draw a parallel between our formulation and a similar one which exists in the purely relativistic case, or equivalently in the acoustic limit. This is intended on the one hand to show how our results smoothly reduce to traditional relativistic results as well as building a bridge between the optical and gravitational dynamics beyond the acoustic approximation.


The BdG equation \eqref{E:BdG} is a set of first order in time PDEs, one for $\psi$ and another for $\overline{\psi}$ which are coupled. They can be re-arranged without approximation into the decoupled 2nd order in time PDE
\begin{align}
&\left(\partial_t+v\cdot \nabla+\i\frac{\alpha}{\rho}\nabla\cdot \rho\nabla\right)\frac{1}{\rho}\left(\partial_t+v\cdot \nabla-\i\frac{\alpha}{\rho}\nabla\cdot \rho\nabla\right)\psi \notag \\
&\hspace{30mm}-2\alpha\beta\frac{1}{\rho}\nabla\cdot \rho\nabla \psi=0 \label{E:full_wave}
\end{align}
and its complex conjugate.  Under the appropriate WKBJ approximation, both these PDEs possesses the Bogoliubov dispersion \eqref{E:bogo}. This is very similar to what we did in \eqref{E:signal_idler_eqns} but here we have made no assumptions about the form of the solution $\psi$.

Equation \eqref{E:full_wave}, however, is not a wave equation \textit{per se} as it stands due to the third term inside each of the brackets. This non-hyperbolic PDE can be approximated by a hyperbolic equation when $\alpha$ is small  while keeping the product $\alpha\beta$ fixed as we mentioned at the end of Sec.~\ref{S:super}.   In this limit \eqref{E:full_wave} can be re-expressed, using \eqref{E:euler1} as
\begin{align}
\left(\partial_t+ \nabla\cdot \mbf{v}\right)\left(\partial_t+\mbf{v}\cdot \nabla\right)\psi-\nabla\cdot c_s^2\nabla \psi=0 \label{E:dAlem},
\end{align}
where $c_s=2\alpha\beta$ is the universal (but position dependent) propagation speed, which is precisely the d'Alembertian describing the dynamics of a massless minimally coupled scalar field propagating in a geometry with metric
\be
ds^2=\left(\frac{\rho}{c_s^2}\right)^2\left[-c_s^2dt^2+\left(d\mbf{x}-\mbf{v}dt\right)^2\right]. \label{E:metric}
\ee

Metrics of the form \eqref{E:metric} are capable of describing a wide range of astrophysically relevant axisymmetric geometrical structures such as black hole event horizons (locations where $v_r^2-c_s^2$ changes sign), ergo-regions (regions where $\mbf{v}^2>c_s^2$) and time dependent cosmological horizons. These structures are the necessary and sufficient geometric ingredients for Hawking radiation, superradiance and cosmological particle production respectively and as such are attractive analogues to simulate in the lab due to absence of experimental verification of these astrophysical phenomena.  

For example, a rotating draining fluid vortex flow $v_r\propto1/r$, $v_\theta\propto 1/r$ has been the subject of much attention in the literature from this perspective \cite{Torres} and is an analogue for a rotating black hole spacetime, possessing both an event horizon and an ergo-region. In optics the rotating spacetime with $v_\theta\propto 1/r$ naturally arises for an optical beam with orbital angular momentum (OAM) (see \cite{Marino} for example). Expanding cosmologies naturally arise in this fashion by switching off the trapping potential of a confined BEC \cite{Eckel}. 

The action for which \eqref{E:dAlem} is the extremal condition is given by
\be
S=\frac{1}{2}\int dt d^nx\, \sqrt{\mbf{g}}\,g^{\mu\nu}\left(\partial_\mu \psi\right)\left(\partial_\nu \overline{\psi}\right)
\ee
which possessed conserved current 
\begin{align}
J^\mu&= \sqrt{g} g^{\mu\nu}\left(\overline{\psi}\partial_\nu\psi-\left(\partial_\nu \overline{\psi}\right)  \psi\right)   \label{E:dAcurr}
\end{align}
and Noether charge
\begin{align}
Q(t)&=\int d^nx \sqrt{\mbf{g}}\,g^{\mu 0}\left(\overline{\psi}\partial_\mu \psi -\text{c.c}\right) \notag \\
&=\int d^nx \, \left(\overline{\psi}\left(\partial_t+\mbf{v}\cdot\nabla\right)\psi-\text{c.c}\right)
\end{align}
which is conserved $dQ/dt=0$ (see \cite{jacobson,Ford} for example). 

As we did in Sec.~\ref{S:trans}, integrating the current density $J^0$ only over a subregion $\mathcal{M}$ with boundary $\partial \mathcal{M}$ results in the `partial charge' $Q_\mathcal{M}$ which will increase when negative norm component modes cross $\partial \mathcal{M}$. This is precisely what happens during superradiance: a positive charge wavepacket propagates towards the inhomogeneous rotational flow, scatters from it into a positive charge reflected mode and a negative charge transmitted packet (see Fig.~\ref{F:2}). 
Of course the sum of these two components is always equal to the initial charge (since total charge is conserved) the charge of the reflected mode will be greater than the initial incident mode. 


Looking closely at the currents in \eqref{E:dcurr}, it is not clear what becomes of them in the acoustic limit $\alpha\rightarrow 0$ with $\alpha\beta=$const. However, their raw forms
\begin{align}
J^0&=U\left(\partial_t-\i m v-\i\alpha \nabla_m^2\right)\overline{V} \notag \\
&\hspace{20mm}-\overline{V}\left(\partial_t+\i m v-\i\alpha \nabla_m^2\right)U 
\end{align}
and 
\be
J^1=-2\alpha\beta \left[U\left(\partial_x\overline{V}\right)-\overline{V}\left(\partial_x U\right)\right] +\mathcal{O}(\alpha)
\ee
provide more insight. We see that the current simply converges to the d'Alembertian current \eqref{E:dAcurr} in the acoustic limit. 

Note also by \eqref{E:modes} we see that in the acoustic limit the positive and negative modes become degenerate being merely conjugates of one another, and from \eqref{E:signal_idler_eqns} we see that both are governed by the same curved spacetime d'Alembertian.

For completeness, the acoustic spacetime associated with our idealised Zel'dovich beam profile in \eqref{S:super} is given by
\be
ds^2\propto -c_s^2dt^2+dr^2+r^2\left(d\theta-\Omega dt\right)^2.
\ee
Such a spacetime can obviously not arise as a solution to dynamical gravitational equations such as the Einstein equations due to the non-differentiability at $r_0$.  However, the spacetime possesses an ergo-region if the rotating region is large enough, $r_0>c_s/\Omega$ since $g_{tt}=-c_s^2+r^2\Omega^2$. While the condition $\omega<m\Omega$ can always be fulfilled for incident wave packets, even for small Zel'dovich cylinders, cylinders with $r_0<c_s/\Omega$ sit inside the turning points of the radial Bessel modes when $\omega<m\Omega$, where they transition to polynomial decay and therefore do not interact strongly with the modes\footnote{This can be estimated by the following argument. Locally, outside the rotating region, a radial plane wave exp$(-\i\omega t+\i m\theta-\i k(r) r)/\sqrt{r}$ has position dependent dispersion $\omega^2\simeq c^2k(r)^2+c^2m^2/r^2$ so that $c^2k(r)^2=\omega^2-c^2m^2/r^2$. If $\omega<m\Omega$ then $c^2k(r)^2<m^2\Omega^2-c^2m^2/r^2$. As this mode approaches the critical radius $r_\text{crit}=c_s/\Omega$ the right hand side of this inequality goes to zero, implying the wavelength goes to infinity and the mode ceases to be oscillatory.  Hence rotating regions smaller than $r_\text{crit}$ cannot superradiate despite satisfying the superradiance condition, in consistency with the spacetime picture which tells us there is no ergo-region for such small cylinders.   }.

\section{Conclusion}

In summary, we have explored the problem of superradiant scattering in the time domain using conserved Noether currents. This has allowed us to generalise the conservation condition predicting the amplitude of superradiant reflections to account for quantum pressure,  which is always present in BEC and photon fluid analogues while making the limit to the superfluid regime explicit where the analogy with curved spacetime is available. It has also made a general transient description of superradiance possible, which permits arbitrary initial conditions and a wide class of backgrounds from which it is apparent that an event horizon is not necessary to observe over-reflection from a rotating spacetime. This transient formalism will be crucial for future experiments on superradiance in photon fluids, in which practical constraints limit potential superradiance measurements to very short time scales. We plan on conducting a complete numerical study of generalised superradiance from this perspective in future work.



\appendix

%
%
%
%
%
%
%
%
%
%
%
%
%

{\textit{Acknowledgements}}--- D.F. and A.P. acknowledge financial support from the EPSRC (UK, Grant No. EP/M009122/1) and E.U. Horizons 2020 (Marie-Sk\l{}odowska Curie Actions). This work has also received funding from the European Union?s Horizon 2020 research and innovation programme under grant agreement No 820392. C.M. acknowledges studentship funding from EPSRC under CM-CDT Grant No. EP/L015110/1.

\bibliographystyle{utphys}

\bibliography{TheoryToExpBib}

\end{document}